\documentclass[12pt]{article}
\usepackage{theorem}
\title{\textsf{Explicit formula for singular vectors of the Virasoro algebra 
with central charge less than 1}}

\author{Reiho Sakamoto\\
\normalsize{\textit{Department of Physics, Graduate School of Science}} \vspace{-2mm}\\
\normalsize{\textit{University of Tokyo, Hongo, Bunkyo-ku, Tokyo, 113-0022, Japan}}
\vspace{-2mm}\\
\normalsize{\texttt{reiho@monet.phys.s.u-tokyo.ac.jp}}}

\date{September 18, 2004}

\begin{document}
\maketitle

\begin{abstract}
We calculate explicitly the singular vectors of the Virasoro algebra
with the central charge $c\leq 1$.
As a result, we have an infinite sequence of the singular vectors
for each Fock space with given central charge and highest weight,
and all its elements can be written in terms of the Jack symmetric functions
with rectangular Young diagram.
\end{abstract}

Since the paper of Belavin, Polyakov and Zamolodchikov \cite{BPZ},
singular vectors of the Virasoro algebra attract a lot of attention
because of its relationship with correlation functions.
Many attempts were done to make the explicit expression of these vectors
in the Verma module (see for example {[}2-4{]}),
however it remains to be unclear what is the structure of these vectors.
On the other hand, singular vectors on the Fock space are known to have
relationships with symmetric functions.
Goldstone, Wakimoto and H. Yamada showed that when the central charge $c$
is equal to $1$, then the singular vectors of smallest degree 
are proportional to the Schur symmetric functions 
with rectangular Young diagram \cite{Go,WY}.
Using the actions of the Virasoro generators on the Schur symmetric functions,
\cite{CDK} generalized this result to the case $c<1$ and expressed the
singular vectors as a sum of the Schur symmetric functions.
Mimachi and Y. Yamada extended these results to the general value of $c$
and showed that the singular vector of the Fock space $\mathcal{F}_{A_{r+1,s+1}}$ 
(see below for definition) with degree $rs$
is proportional to the Jack symmetric function with rectangular
Young diagram $(s^r)$ \cite{TK,MY}.
However, as is well known, there are infinitely many singular vectors
in each Fock space $\mathcal{F}_{A_{r+1,s+1}}$ if $c\leq 1$,
and all above results determine only the singular vectors of smallest degree.
Recently, \cite{SSAFR} demonstrated that the set of all the bosonized
Jack symmetric functions forms a natural basis of the Fock space,
and conjectured explicitly the actions of the
Virasoro generators on this Jack basis.
In this paper, applying these results, we determine the explicit expression
of all the singular vectors of the Fock space $\mathcal{F}_{A_{r+1,s+1}}$
when central charge is less than 1.
As a result, we found that all the singular vectors are proportional to the
Jack symmetric functions with rectangular Young diagram,
which indicates further deep relationships between the Jack symmetric functions
and the representations of the Virasoro algebra.

\bigskip

The set-up is as follows.
Denote the Virasoro generators as usual by $L_n$ $(n\in \mathbf{Z})$
and the central charge as $c$, with commutation relations
\begin{equation}
{[} L_n ,L_m {]} =(n-m)L_{n+m}+\frac{n(n^2 -1)}{12}\, c\, \delta_{n+m,0}.
\end{equation}
As a representation space, we take the Fock space $\mathcal{F}_A$ defined by
\begin{equation}
\mathcal{F}_A :=\mathbf{C}{[}a_{-1},a_{-2},a_{-3},\cdots {]}|A\rangle ,
\end{equation}
where we use the bosonic operators ${[}a_n ,a_m {]}=n\, \delta _{n+m,0}$
$(n,m\in\mathbf{Z})$, and a vacuum vector $|A\rangle $ defined by
\begin{equation}
a_0 |A\rangle =A|A\rangle , \quad a_n |A\rangle =0\,\,
(n\in \mathbf{Z}_{>0} )
\end{equation}
Then the bosonic representation of $L_n$'s is given as follows 
--- the Feigin-Fuchs representation \cite{FF1},
\begin{equation}
L_n=\frac{1}{2}\sum_{k\in\mathbf{Z}} :a_{n-k}a_{k}:
-A_{1,1}(n+1)a_n ,
\end{equation}
here we use the notation
\begin{equation}
A_{r,s}=\frac{1}{\sqrt{2}}\left(r\sqrt{\beta}
-\frac{s}{\sqrt{\beta}}\right) .
\end{equation}
This $\beta $ is related to the central charge $c$ as
\begin{equation}
c=1-\frac{6\, (1-\beta)^2}{\beta} .
\end{equation}
Note that by the universality of the Verma module, we have a map
from the Verma module with highest weight 
\begin{equation}
h_{r,s}=\frac{(r\beta -s)^2-(\beta -1)^2}{4\beta},
\end{equation}
to the Fock space $\mathcal{F}_{A_{r+1,s+1}}$ 
(we have $L_0 |A_{r+1,s+1}\rangle =h_{r,s}|A_{r+1,s+1}\rangle$).

Another ingredient of this study is the Jack symmetric function.
Consider the Hamiltonian
\begin{equation}
H_{\beta}:=\sum_{i=1}^{N}\left( x_i \frac{\partial}{\partial x_i}
\right)^2 +\beta\, \sum_{i<j} \frac{x_i +x_j}{x_i -x_j}
\left(x_i\frac{\partial}{\partial x_i}-x_j\frac{\partial}{\partial x_j}
\right) ,
\end{equation}
where $N$ is a number of variables.
Then the Schr\"{o}dinger equation
\begin{eqnarray}
H_{\beta} J_\lambda (x_1,x_2,\cdots ,x_N)&=&
\epsilon_\lambda J_\lambda (x_1,x_2,\cdots ,x_N) ,\\
\epsilon_\lambda &=&\sum_{i=1}^{N}\left(
\lambda_i^2 +\beta\, (N+1-2i)\lambda_i \right),
\end{eqnarray}
is known to be exactly solvable and its wave functions are called the 
Jack symmetric polynomials \cite{Jack,St}.
These $J_\lambda$'s are the $N$ variable symmetric polynomials 
parameterized by partitions $\lambda =(\lambda_1,\lambda_2,
\lambda_3,\cdots )$.
Physically, Jack polynomials express the excited states of the Calogero-Sutherland model {[}14-17{]}
with periodic boundary condition.
Ujino, Hikami and Wadati investigated the complete quantum integrability
of these systems \cite{UHW,UWH} (see also \cite{HW}).

Instead of using the coordinates $x_i$,
we can express symmetric polynomials uniquely in terms of power sums 
$p_n :=\sum x_i^n$, if the number of variables is sufficiently large.
After expressing the Jack polynomials by power sums, 
we can take the number of variables as infinity.
These Jack symmetric polynomials with infinitely many variables
are called Jack symmetric functions.

To characterize Jack symmetric functions by orthogonality, we define the inner product on 
the space of symmetric functions as
\begin{eqnarray}
\langle p_\lambda , p_\mu \rangle &:=& \delta_{\lambda ,\mu}
z_\lambda \beta^{-l(\lambda )}, \\
p_\lambda &:=&\prod_{i=1}^{l(\lambda )}p_{\lambda _i},\\
z_\lambda &:=&\prod_{i} i^{m_i}\cdot m_i ! .
\end{eqnarray}
Here, $l(\lambda )$ is a length of partition $\lambda$ and $m_i$'s
are the multiplicity of $i$ in partition $\lambda$, i.e. we have
$\lambda =(\cdots 3^{m_3},2^{m_2},1^{m_1})$.
We also denote the weight of partition $\lambda$ as $|\lambda |$
and define the coordinate of partition like a matrix, i.e.
if the box is on row $i$ and column $j$, then the coordinate is $(i,j)$.
Using this coordinate, define the arm length $a(s)$ and leg length $l(s)$ of the box 
$s=(i,j)\in \lambda$ by
\begin{equation}
a(s):= \lambda _i-j,\quad l(s):=\lambda_j^{'}-i,
\end{equation}
where $\lambda_j^{'}$ is the $j$-th element of the transposed partition $\lambda^{'}$.

Then our normalization $J_\lambda$ is characterized by its norm \cite{St}
\begin{equation}
\langle J_\lambda ,J_\mu\rangle =\delta_{\lambda ,\mu }
\frac{1}{\beta^{2|\lambda |}}\prod_{s\in\lambda }
\left( a(s)+(l(s)+1)\beta \right) ((a(s)+1)+l(s)\beta ) ,
\end{equation}
especially we set the coefficient in front of $p_1^{|\lambda |}$ to be unity.
See \cite{Mac} for rigorous discussion about uniqueness.
As an example, we give a list of Jack symmetric function with degree 3
(here we set $\alpha :=1/\beta$),
\begin{eqnarray*}
J\unitlength 4pt
\begin{picture}(3,1)
\multiput(0,0)(1,0){4}{\line(0,1){1}}
\multiput(0,0)(0,1){2}{\line(1,0){3}}
\end{picture}
&=& 2\alpha^2 p_3+3\alpha p_2 p_1+p_1^3 ,\\
J\unitlength 4pt
\begin{picture}(3,2)(0,1)
\multiput(0,0)(1,0){2}{\line(0,1){2}}
\put(0,0){\line(1,0){1}}
\multiput(0,1)(0,1){2}{\line(1,0){2}}
\put(2,1){\line(0,1){1}}
\end{picture}
&=& -\alpha p_3+(\alpha -1)p_2 p_1+p_1^3 ,\\
J\unitlength 4pt
\begin{picture}(3,3)(0,2)
\multiput(0,0)(1,0){2}{\line(0,1){3}}
\multiput(0,0)(0,1){4}{\line(1,0){1}}
\end{picture}
&=& 2p_3-3p_2 p_1+p_1^3 .
\end{eqnarray*}

{}From this characterization, we can identify the Jack symmetric function
and element of the Fock space \cite{AMOS1,AMOS2}.
To do this, we use the following two identifications,
\begin{eqnarray}
p_n &\longleftrightarrow & \sqrt{\frac{2}{\beta}}a_{-n}|a\rangle ,\\
p_n &\longleftrightarrow & \langle A|a_n \frac{1}{\sqrt{2\beta}} .
\end{eqnarray}
In \cite{SSAFR}, it was proposed that we should take the basis of the 
Fock space as a set of all the bosonized Jack symmetric functions,
\begin{equation}
  \mathcal{F}_A =\mathrm{span}\Bigl\{ J_{\lambda} |A\rangle \Bigl| \mathrm{any\,
  partition}\, \lambda \Bigl\} .
\end{equation}

Since \cite{SSAFR} deals with only the leftwards action of $L_n$ operators on
this Jack basis, we need to rewrite the formula for rightwards action.
In subsequent discussion, 
we write $J_\lambda |A_{r+1,s+1}\rangle =|J_\lambda \rangle$ for simplicity,
and also denote the action of $L_n$ as
\begin{eqnarray}
\langle J_\lambda |L_n &=&\sum_{|\mu |-|\lambda |=n}\langle J_\mu |c_{\lambda\mu},\\
L_n |J_\mu\rangle &=& \sum_{|\mu |-|\lambda |=n}d_{\mu\lambda}|J_\lambda\rangle .
\end{eqnarray}
Note that $c_{\lambda\mu}$'s can be determined from the formula in \cite{SSAFR}.
Then from the identity
\begin{eqnarray}
 & & (\langle J_\lambda |L_n)|J_\mu\rangle =c_{\lambda\mu}
     \langle J_\mu |J_\mu \rangle  \nonumber\\
 &=& \langle J_\lambda |(L_n| J_\mu\rangle )=d_{\mu\lambda}
     \langle J_\lambda |J_\lambda \rangle ,
\end{eqnarray}
we have the following relation,
\begin{equation}
d_{\mu\lambda}=\frac{\langle J_\mu |J_\mu \rangle}{\langle J_\lambda |J_\lambda\rangle}
c_{\lambda\mu} .
\end{equation}

First we consider the action of $L_1$ on Jack symmetric function 
with rectangular partition.
In this case, we have
\begin{eqnarray}
c_{(n^{m-1},n-1),(n^m)} &=&
c\unitlength 4pt
\begin{picture}(9,3)(-0.5,2.5)
\put(0,0){\line(1,0){3}}
\put(0,0){\line(0,1){3}}
\put(3,0){\line(0,1){1}}
\put(3,1){\line(1,0){1}}
\put(4,1){\line(0,1){2}}
\put(0,3){\line(1,0){4}}
%%%%%%%%%%%%%%%%%%%%%%%%%%%%%%%%%%%%%
\multiput(5,0)(0,3){2}{\line(1,0){4}}
\multiput(5,0)(4,0){2}{\line(0,1){3}}
\end{picture}\nonumber\\
 &=&\sqrt{2\beta}\frac{1\cdot\beta}{(1+(m-1)\beta)((n-1)+\beta)}A_{r-m,s-n},\\
\frac{\langle J_{(n^m)}|J_{(n^m)}\rangle}
{\langle J_{(n^{m-1},n-1)}|J_{(n^{m-1},n-1)} \rangle}&=&
\frac{1}{\beta ^2}\cdot
\frac{(0+m\beta )(1+(m-1)\beta)((n-1)+\beta)(n+0\beta )}{(0+\beta)(1+0\beta )}.
\end{eqnarray}
Note that the leftward action $\langle J_\lambda |L_1$ is amount to the addition of one box
to the partition $\lambda$.
Then from the orthogonality of the Jack symmetric functions,
we conclude that the rightward action $L_1 |J_\lambda \rangle$ amount to 
subtract one box from the partition $\lambda$.
Thus we obtain the formula for $J_{(n^m)}$ as
\begin{eqnarray}
L_1 |J_{(n^m)}\rangle&=&c_{(n^{m-1},n-1),(n^m)}\,
\frac{\langle J_{(n^m)}|J_{(n^m)}\rangle}
{\langle J_{(n^{m-1},n-1)}|J_{(n^{m-1},n-1)} \rangle}
|J_{(n^{m-1},n-1)}\rangle\nonumber\\
 &=&\sqrt{\frac{2}{\beta}}\cdot \frac{1}{\beta}\, mn\beta\,
A_{r-m,s-n}|J_{(n^{m-1},n-1)}\rangle .\label{L1}
\end{eqnarray}

Next we consider the action of $L_2$.
In this case, we have the following equalities,
\begin{eqnarray}
c_{(n^{m-1},n-2),(n^m)}&=&c
\unitlength 4pt
\begin{picture}(4,1)(-0.5,3.5)
\put(0,1){\line(1,0){2}}
\put(0,1){\line(0,1){3}}
\put(2,1){\line(0,1){1}}
\put(2,2){\line(1,0){2}}
\put(0,4){\line(1,0){4}}
\put(4,2){\line(0,1){2}}
%%%%%%%%%%%%%%%%%%%%%%%%%
\multiput(5,1)(4,0){2}{\line(0,1){3}}
\multiput(5,1)(0,3){2}{\line(1,0){4}}
\end{picture}\nonumber\\
 &=&\sqrt{2\beta}\,\beta\,
\frac{2\,\beta}{(2+(m-1)\beta)((n-2)+\beta)}\nonumber\\
 & &\times\frac{1}{(1+(m-1)\beta)((n-1)+\beta)}A_{r-m,s-n},\\
\frac{\langle J_{(n^m)}|J_{(n^m)} \rangle}
{\langle J_{(n^{m-1},n-2)}|J_{(n^{m-1},n-2)} \rangle}&=&
\frac{1}{\beta^4}\cdot\frac{1}{(1+\beta )(2+0\beta )(0+\beta )(1+0\beta )}\\
 & &\times(1+m\beta )(2+(m-1)\beta)(0+m\beta )(1+(m-1)\beta )\nonumber\\
 & &\times((n-1)+\beta )(n+0\beta )((n-2)+\beta )((n-1)+0\beta ),\nonumber\\
c_{(n^{m-2},(n-1)^2),(n^m)}&=&c
\unitlength 4pt
\begin{picture}(9,3)(-0.5,2.5)
\put(0,0){\line(1,0){3}}
\put(0,0){\line(0,1){3}}
\put(3,0){\line(0,1){2}}
\put(3,2){\line(1,0){1}}
\put(4,2){\line(0,1){1}}
\put(0,3){\line(1,0){4}}
%%%%%%%%%%%%%%%%%%%%%%%%%%%
\multiput(5,0)(4,0){2}{\line(0,1){3}}
\multiput(5,0)(0,3){2}{\line(1,0){4}}
\end{picture}\nonumber\\
 &=&-\sqrt{2\beta}\frac{1\cdot 2\beta}{(1+(m-2)\beta )((n-1)+2\beta )}\nonumber\\
 & &\times\frac{\beta}{(1+(m-1)\beta)((n-1)+\beta )}\, A_{r-m,s-n},\\
\frac{\langle J_{(n^m)}|J_{(n^m)}\rangle}
{\langle J_{(n^{m-2},(n-1)^2)}|J_{(n^{m-2},(n-1)^2)} \rangle}
&=&\frac{1}{\beta^4}\cdot\frac{1}{(0+2\beta )(1+\beta)(0+\beta )(1+0\beta )}\\
 & &\times(0+m\beta )(1+(m-1)\beta )(0+(m-1)\beta )(1+(m-2)\beta )\nonumber\\
 & &\times((n-1)+2\beta )(n+\beta )((n-1)+\beta )(n+0\beta ).\nonumber
\end{eqnarray}
Combining these equalities, we have
\begin{eqnarray}
L_2 |J_{(n^m)}\rangle &=&c_{(n^{m-1},n-2),(n^m)}\, 
\frac{\langle J_{(n^m)}|J_{(n^m)} \rangle}
{\langle J_{(n^{m-1},n-2)}|J_{(n^{m-1},n-2)} \rangle}\,
|J_{(n^{m-1},n-2)}\rangle\nonumber\\
 & &+\, c_{(n^{m-2},(n-1)^2),(n^m)}\,
\frac{\langle J_{(n^m)}|J_{(n^m)}\rangle}
{\langle J_{(n^{m-2},(n-1)^2)}|J_{(n^{m-2},(n-1)^2)} \rangle}\,
|J_{(n^{m-2},(n-1)^2)}\rangle\nonumber\\
 &=&\sqrt{\frac{2}{\beta}}\cdot\frac{1}{\beta^2}\left(
mn\beta\cdot\frac{(n-1)(1+m\beta )}{(1+\beta )}\, A_{r-m,s-n}
|J_{(n^{m-1},n-2)} \rangle\right. \nonumber\\
 & &\left. \qquad\quad\,\,\,\, -mn\beta\cdot
\frac{(m-1)\beta(n+\beta)}{(1+\beta )}\, A_{r-m,s-n}
|J_{(n^{m-2},(n-1)^2)} \rangle \right) .\label{L2}
\end{eqnarray}

{}From now on, we restrict the value of $\beta$ as follows,
\begin{equation}
\beta =\frac{p}{q}>0,
\end{equation}
where $p$ and $q$ are mutually prime positive integers.
When we use this value of $\beta$, the central charge of the Virasoro
algebra satisfies $c\leq 1$.
If we rewrite the definition of $A_{r,s}$ in this case as
\begin{equation}
A_{r,s}=\frac{1}{\sqrt{2\beta}}\left( \frac{p r}{q}-s\right),
\end{equation}
we have the equality
\begin{equation}
A_{m-(m+aq),n-(n+ap)}=A_{-aq,-ap}=0,\quad a\in\mathbf{Z}_{>0}.\label{zero}
\end{equation}
Combining this simple equality with equations (\ref{L1}) and (\ref{L2}),
we obtain the following remarkable result.

\textbf{Theorem}
\textit{If $\beta =p/q>0$, then the Fock space $\mathcal{F}_{A_{m+1,n+1}}$
$(m,n\in\mathbf{Z}_{>0})$ have the following infinite sequence of singular vectors;}
\begin{equation}
|J_{(n^m)}\rangle ,|J_{(n+p)^{(m+q)}}\rangle 
,|J_{(n+2p)^{(m+2q)}}\rangle ,\cdots ,|J_{(n+ap)^{(m+aq)}}\rangle ,\cdots . \label{seq}
\end{equation}

To verify this, we consider the action of $L_1$ and $L_2$ on $|J_{(n+ap)^{(m+aq)}}\rangle$.
Then from equations (\ref{L1}) and (\ref{L2}), we have the term
$A_{m-(m+aq),n-(n+ap)}$, which is equal to $0$ by (\ref{zero}).

We remark that the first element of the sequence (\ref{seq}) coincide with the result given in \cite{MY}
(see also \cite{Do,FF2} for the structure of the Verma modules).
We also remark that in (\ref{seq}), we do not take care of the precise normalization.

\textbf{Example} We consider the case $\beta =3/4$.
This amount to $c=1/2$ to which the critical Ising model corresponds.
If we choose the highest weight as $A_{1+1,2+1}$,
then we have a sequence of singular vectors
\begin{equation}
|J_{(2)}\rangle ,|J_{(5^5)}\rangle ,|J_{(8^9)}\rangle 
,|J_{(11^{13})}\rangle ,\cdots .
\end{equation}

\bigskip

\textbf{Acknowledgments:}
The author is grateful to Miki Wadati for helpful discussions and careful reading of the manuscript.
The author is also grateful to Boris Feigin for valuable comments on this manuscript. 

\pagebreak

\end{document}